\documentstyle[psfig,12pt]{article}
\def\nl{\paragraph\noindent}
\def\nn{\nonumber}
\def\ds{\displaystyle }

\def\Journal#1#2#3#4{{#1} {\bf #2} (#3) #4.}
\def\NCA{\em Nuovo Cimento}

\def\NPB{{\em Nucl. Phys. B}}
\def\PLB{{\em Phys. Lett. B}}
\def\PRL{\em Phys. Rev. Lett.}
\def\PR{{\em Phys. Rev.}}

\title{ Unitarized ChPT Amplitudes \\
and  Crossing Symmetry  Violation}

\author{Isabela P. Cavalcante \ and
J. S\'a Borges \\  \\
 Universidade do Estado do Rio de Janeiro\\ 
Rio de Janeiro, Brazil}
\date{}

\begin{document}
\maketitle

\begin{abstract}
 
 Pion-pion scattering amplitude obtained from one-loop Chiral
Perturbation Theory (ChPT)  is crossing symmetric, however the
corresponding partial-wave  amplitudes do not respect exact
unitarity relation. There are different approaches to get 
unitarized results from ChPT. Here we consider  the inverse
amplitude method (IAM) and, using the Roskies relations,  we measure the amount of crossing
symmetry violation when IAM is used  in order to fit pion-pion
phase-shifts  to experimental data in the resonance region.
We also show the unitarity violation of the crossing symmetric ChPT
amplitude  with its  parameters fixed in order to fit to 
experimental phase-shifts. 
\end{abstract}

\nl

\section{Introduction}

Even though Quantum Chromodynamics (QCD) has achieved a great success
in describing strong interactions, low energy hadron physics must
still be modeled phenomenologically.  
A great theoretical improvement was made by means of the method of
ChPT \cite{Leu}, which is an effective theory
derived from the basis of QCD.  The method consists of writing down chiral
Lagrangians for the physical processes and  uses the conventional
technique of the field theory for the calculations. 

Here we will focus on pion-pion
scattering. For this process, 
the   ChPT leading  contribution  (tree graphs) is of second order in the
momenta $p$ of the external pions and coincides with   Weinberg 
result from  current algebra \cite{Wei}.
 The corrections come from  loop diagrams
whose  vertices are of order $p^2$ and include a free-parameter polynomial 
part related to 
tree diagrams of order $p^4$; these
parameters have to be  obtained phenomenologically.
At  one-loop level the method  yields a total
amplitude that respects exact crossing symmetry, however, the corresponding 
partial-waves satisfy only approximate elastic unitarity.

This violation is more severe at higher energies, so that it is not possible to
reproduce resonant states, which are one of the most relevant features
of the strong interacting regime. This is not a new issue in
literature and
many different methods have been
proposed to improve this behaviour.
Here we  consider the 
inverse amplitude method (IAM) \cite{dob}, that allows 
one to access the resonance region for pion-pion scattering by
fixing  two parameters, but violates crossing symmetry.  

 In the present exercise, our goal is to quantify this violation of
crossing symmetry,  what we do  by calculating the deviations from 
 the Roskies relations \cite{Ro}.
Our work is presented as follows.
In section \ref{sec:iam} we write the ChPT amplitude for pion-pion
scattering and we construct IAM partial-waves. 
We introduce  a correction to get  rid of sub-threshold poles 
by slightly shifting the original Adler zeros of the
 leading amplitudes.
In section \ref{sec:roskies} we display the so called 
Roskies relations, which follow from the requirement of exact crossing symmetry and involve
integrals of the partial-wave amplitudes in the region $\, 0 < s < 4
m_\pi^2$.  
We measured the crossing symmetry violation of the 
   IAM  amplitudes. We obtained that  some Roskies relations are
violated at 30\% \ level. 

The compromise between crossing symmetry violation and approximate
unitary amplitudes can be established by computing unitarity violation
of the crossing symmetric ChPT amplitude, with parameters fixed in
order to reproduce 
experimental phase-shifts. This is done in section \ref{sec:con}
that also presents a summary of the main results.

\section {Chiral perturbation theory  and the IAM}
\label{sec:iam}

In the case of pion-pion scattering, crossing symmetry implies that
there is just one amplitude 
describing the three total isospin channels of the process. Using ChPT
at the one-loop level and considering only the most relevant low
energy constants,  the amplitude can be decomposed as 
$$A(s,t,u)= A^{ca}(s,t,u) + B(s,t,u) + C(s,t,u), \quad
{\hbox{where}}$$
\begin{eqnarray*}
f_\pi^2 \, A^{ca}(s,t,u) &=& s - m_\pi^2 \, , \\
f_\pi^4  \, B(s,t,u) &=& 
\frac 1 6 \,\left (4\,\left (s-1/2\,{m_\pi}^{2}\right )^{2}-\left (s-2\,{
m}^{2}\right )^{2}\right )\bar J(s) \\
&+& \left[ \frac 1 {12}\, \left (3\,\left (t-2\,{m_\pi}^{2}
\right )^{2}+\left (s-u\right )\left (t-4\,{m_\pi}^{2}\right )\right
)\bar J(t) + (t \leftrightarrow u) \right]  \, , \\
f_\pi^4 \, C(s,t,u) &=&
\lambda_{{1}}\left (s-2\,{m_\pi}^{
2}\right )^{2}+\lambda_{{2}}\left[ \left (t-2\,{m_\pi}^{2}\right )^{2}
 + (t \leftrightarrow u) 
\right] \, .
\end{eqnarray*}
The isospin defined amplitudes  $T_I$\, for $I\, =\,  0, 1\,\,
{\hbox{and}}\, \,  2$ are
\begin{eqnarray*}
T_0 (s,t) &=& 3 A(s,t,u)\, +\, A(t,s,u)\, + \, A(u,t,s) \, , \\
T_1 (s,t) &=&  A(t,s,u)\, - \, A(u,t,s) \, , \\
T_2 (s,t) &=&  A(t,s,u)\, + \, A(u,t,s)\, ,
\end{eqnarray*}
which are
expanded in partial-wave amplitudes, as
$$ T_I(s,t)\, = \, \sum_{\ell} (2 \ell + 1 ) \, t_{\ell\, I}(s) \, P_\ell \, (\cos
\theta), $$
where $P_\ell$ are the Legendre polynomials. In the following we omit
the label $\ell$, because we will just deal with S-wave  ($I=0\, , 2$) and
P-wave  ($I= 1$).

Elastic unitarity implies that,  for\,  $ 16 m_\pi^2 \ge s \ge 4
m_\pi^2$\, , 
$$\mbox{Im} \, t_{I} (s) = \rho (s) |  t_{I} (s) |^2,$$
which can be solved yielding 
\begin{equation}
 t_{I} (s) = \frac 1 {\rho
(s)} e^{i\, \delta_I(s)}\, \sin \delta_I(s),
\label{delta}
\end{equation}
where $ \delta_I(s)$ are the real phase-shifts and
$$\rho(s) = \frac 1 { 16 \pi} \sqrt{\frac{s - 4 m_\pi^2}{s}}$$ 
is the phase space
factor for pion-pion scattering. 
Even for the  ChPT amplitude, that does not respect elastic unitarity
 constraint, the definition  
\begin{equation}
 \delta_I(s) = \arctan \frac {\mbox{Im}\, t_I(s)}{\mbox{Re} \, t_I(s)},
\label{upca}
\end{equation}
will be used, in section \ref{sec:con}.

 At one-loop level, that is, up to order $p^4$ in the chiral expansion,
the resulting  
ChPT amplitudes for  isospin $I = 0, 2$ ($\ell = 0$, S-wave)  and $I=1$
($\ell = 1$,  P- wave) can be expanded as
\begin{equation}
 t_I (s) = t_I^{ca}(s) + t_I^{ca\ ^2}(s)\,  \bar J(s) +
t_I^{left}(s) +p_I(s),
\label{chpt}
\end{equation}
where $t_{I}^{ca}$
are the (real) Weinberg amplitudes, namely,
\begin{eqnarray*} 
f_\pi^2 \, t_0^{ca}(s) = 2 s - m_\pi^2, \quad f_\pi^2\, t_1^{ca}(s) =
\frac 1 3\, (
s - m_\pi^2),\quad  f_\pi^2\, t_2^{ca}(s) =  2 \, m_\pi^2
- s,
\end{eqnarray*}
\  $t_I^{left}$ are  the parts that bear
the left-hand cuts, namely, 
\begin{eqnarray*}
f_\pi^4 \pi^2  \, t_0^{left}(s)&=&\frac 1 {12} \frac {m_\pi^4}{s-4
m_\pi^2} ( 6 s - 25 m_\pi^2)\, L(s)^2 - \frac 1 {72 \rho(s)} ( 7 s^2 -
40 m_\pi^2 s + 75 m_\pi^4) L(s)  \\ 
&+& \frac 1 {864} ( 95 s^2 - 658 m_\pi^2
s + 1454 m_\pi^4),\\
f_\pi^4 \pi^2  \, t_2^{left}(s) &=& \frac {-1} {12} \frac {m_\pi^4}{ s  - 4
m_\pi^2} ( 3 s + m_\pi^2)\, L(s)^2 - \frac 1 {144 \rho(s)} ( 11 s^2 -
32 m_\pi^2 s + 6 m_\pi^4) L(s)\\ &+& \frac 1 {1728} ( 157 s^2 - 494 m_\pi^2
s + 580 m_\pi^4), \\
\lefteqn{f_\pi^4 \pi^2 ( s - 4 m_\pi^2)\,  t_1^{left}(s) \, = \, \frac 1 {12} \frac {m_\pi^4}{s-4
m_\pi^2} ( 3 s^2  - 13 m_\pi^2 s - 6 m_\pi^4)\, L(s)^2 } \hspace*{2cm} \\ 
&+& \frac 1 {144 \rho(s)} (  s^3 -
16 m_\pi^2 s^2 + 72 m_\pi^4 s -36 m_\pi^6) L(s) \\ &-& \frac 1 {864} ( 7
s^3 - 71 m_\pi^2  s^2 + 427  m_\pi^4 s - 840 m_\pi^6),\quad
{\hbox{with}}
\end{eqnarray*}
$$\bar J (s) = \frac 1 {8 \pi^2} -\frac 2 {\pi}\,  \rho (s)\, \,  L(s)
+ I\, 
\rho (s) \, , \quad  \quad
L(s)\,  =\,  \ln \, \frac{\sqrt{s-4 m_\pi^2} + \sqrt{s}}{2 m_\pi} \, ,$$ 
 and $ p_I(s)$\ are two free parameter polynomials,
given by
\begin{eqnarray*}
f_\pi^4 \, p_0 (s) &=& 
\frac 1 3 \, 
\left (11\,{s}^{2}-40\,s{m_\pi}^{2}+44\,{m_\pi}^{4}\right
)\lambda_{{1}}+\frac 1 3 \left (14\,{s}^{2}-40\,s{m_\pi}^{2}+56\,{m_\pi}^{4}\right
)\lambda_{{2}}, \\
f_\pi^4 \, p_1 (s) &=& \frac 1 3 s \left(s - 4 m_\pi^2 \right) \left( \lambda_2
- \lambda_1\right) \, ,\\
f_\pi^4 \, p_2 (s) &=& 
\frac 2 3 \,
\left ({s}^{2}-2\,s{m_\pi}^{2}+4\,{m_\pi}^{4}\right )\lambda_{{1}}+
\frac 2 3 \, \left
(4\,{s}^{2}-14\,s{m_\pi}^{2}+16\,{m_\pi}^{4}\right )\lambda_{{2}} \, .
\end{eqnarray*}

\begin{figure}[p]
 \centerline{\psfig{figure=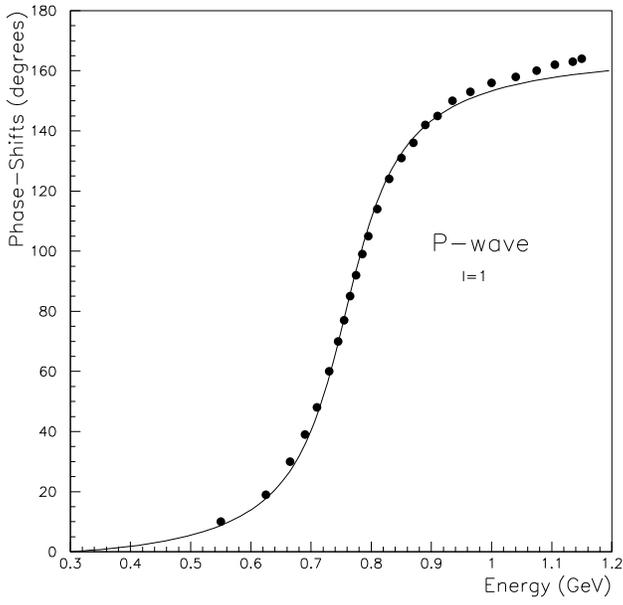,height=9.0cm}\hspace{0.5cm}
\psfig{figure=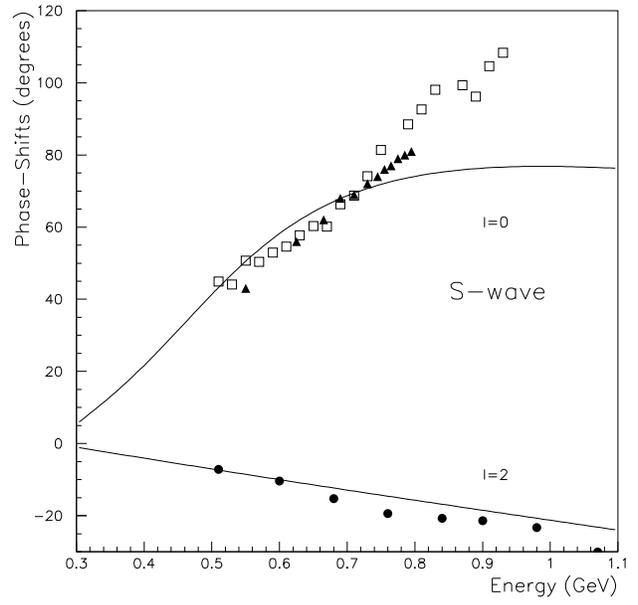,height=9.0cm}}
\caption{Results from fits of IAM amplitudes 
to P- and S-wave phase-shifts, in
degrees, as functions of cms energy, in GeV. Experimental data for
P-wave are from
Ref. \cite{pro};  for S-wave, from Refs. \cite{pro,estra,lost}.}
\label{iam_fits}
\end{figure}

If one wants to describe a resonant amplitude, one may wish to use
Pad\'e approximants,  as e.g. advocated in \cite{dob}. It amounts to writing the inverse of the  partial-wave. Thus,  instead of
the exact ChPT result $t_I$, we use a modified 
amplitude 
\begin{equation}
\tilde t_I (s) = \frac {t_I^{ca}(s)}{ 1 -  \left( t_I^{ca\ ^2}(s)\,  \bar J(s) +
t_I^{left} (s) + p_I(s)\right)/t_I^{ca}(s)}, \quad I=0,1\ {\hbox{and}}\ 2.
\label{ttil}
\end{equation}
Our strategy was  to choose the parameters $\lambda_1$\, and $\lambda_2$
in order to fit S- and P-waves above to the experimental phase-shifts,
by using the definition (\ref{delta}).
We show in Fig. \ref{iam_fits} the resulting phase-shifts
corresponding to the parameters $\lambda_1 = -0.00345$ and $\lambda_2
= 0.01125$. 
As mentioned in the introduction, there was a problem concerning
S-waves, namely that they were singular  
 at some sub-threshold value for $s$, where  the correction becomes equal to
$t^{ca}$. 
Singularities occur in S-wave sub-threshold amplitudes 
at $s_0\simeq 0.64 \, m_\pi^2$, for $I=0$, and at
$s_2 \simeq 1.95 \, m_\pi^2$,
for $I=2$. Those values are  close to the ones where
$t_0^{ca}$ and $t_2^{ca}$ actually vanish. 

In order to get rid of those singularities,  we performed an extra
correction, thus obtaining a new partial-wave amplitude, denoted by \,
$\tilde t_I^{(n)}$,
\begin{equation}
\tilde t_I^{(n)} (s) = \frac {(s-s_I)/f_\pi^2}{ 1 -  \left( t_I^{ca\ ^2}(s)\,  \bar J(s) +
t_I^{left} (s) + p_I(s)\right)/t_I^{ca}(s)}, \quad I=0\ {\hbox{and}}\ 2.
\label{tn}
\end{equation}
The
new formula slightly 
violates unitarity as can be measured by evaluating the quantity
$$ Y^{IAM} = \frac{ \mbox{Im}\,  \tilde t_I^{(n) \, -1} - \rho}{
\mbox{Im}\, \tilde t_I^{(n) \, -1} + \rho} \, ,$$
that is smaller than 2\% in the energy range considered. One notices
that the fits are not modified due to this correction, according to
the phase-shift definition.

\section{Crossing symmetry violation}
\label{sec:roskies}

As explained in the introduction,  IAM allows one to access the
resonance region for pion-pion scattering. 
On the other hand, the corresponding partial-waves do not respect
crossing symmetry and we would like to  quantify that 
violation.
Crossing symmetry imposes constraints between the integrals 
of some combinations of partial-wave amplitudes,  known as
Roskies relations \cite{Ro}. Let us define
$$
\begin{array}{ll}
A_1 = 2 \, \ds \int_0^{4 m_\pi^2} \, f_0 \, ds \, ,   
&B_1 = 5 \, \ds \int_0^{4 m_\pi^2} \,
f_2 \, ds \, , \\[5 mm]
A_2 = \ds \int_0^{4 m_\pi^2}  (3 s - 4 m_\pi^2)  \,
f_0 \, ds \, , &
B_2 = - 2 \, \ds \int_0^{4 m_\pi^2}  (3 s - 4 m_\pi^2) \,
f_2 \, ds \, , \\[5 mm]
A_3 = \ds \int_0^{4 m_\pi^2}  (3 s - 4 m_\pi^2)  \,
f_0 \, ds \, ,  &
B_3 = 2 \, \ds \int_0^{4 m_\pi^2}  \,
f_1 \, ds \, ,  \\[5 mm]
A_4 = \ds \int_0^{4 m_\pi^2}  (3 s - 4 m_\pi^2) \,
 (2 \, f_0 - 5 \, f_2 ) \, ds \, , &
B_4 = 9 \, \ds \int_0^{4 m_\pi^2} 
\, f_1\, ds  \, ,\\[5 mm]
A_5 =\int_0^{4 m_\pi^2}  (10 s^2 - 32 s m_\pi^2 +16
m_\pi^4) \, (2 \, f_0 - 5 \, f_2 ) \, ds \, , &
B_5 = - 6 \,\int_0^{4 m_\pi^2}
\  (5 s - 4 m_\pi^2)  \, f_1\, ds \, ,  
\end{array}
$$
\begin{eqnarray} 
\vspace*{-7 mm}
\lefteqn{A_6 =\int_0^{4 m_\pi^2}  (35 s^3 - 180 s^2 m_\pi^2 +
240 s m_\pi^4 -64 m_\pi^6)  \, (2 \, f_0 - 5 \, f_2 ) \, ds \,
,} \label{ints}   \\
&\hspace*{6.5 cm}&
B_6 = 15 \, \int_0^{4 m_\pi^2} 
\  (21 s^2 - 48 s m_\pi^2 +16 m_\pi^4) \,  f_1\, ds  \, ,\nn
\end{eqnarray}
where $f_{0,2} = (s-4 m_\pi^2 ) t_{0,2}$\ and $f_1 = (s-4 m_\pi^2 )^2 t_1$.
Crossing symmetric amplitudes
must satisfy  $$A_i = B_i \, ,$$
for $i$ from 1 to 6. In order to quantify the amount of violation of these
relations within the IAM, we evaluated the  ratio
\begin{equation}
V_i = \frac {A_i-B_i}{A_i+B_i} \, .
\label{viola}
\end{equation}
We obtained the values shown in the Table.

\section{Discussion and final remarks}
\label{sec:con}

The  $\cal O$(p$^4$) ChPT pion-pion amplitude is crossing symmetric
but does  not
respect exact elastic unitarity. There are several attempts to
extrapolate the domain of validity of 
ChPT and to access the resonance region for meson-meson
scattering. One of these methods uses the inverse of the amplitude and
fits the two-parameter amplitude to the experimental
data.

In the present exercise we were interested in quantifying the
crossing violation that  this procedure implies.  In order to do that we 
used the Roskies relations and we arrived to very big violations of
crossing symmetry. On the other hand we have shown how to get rid of
sub-threshold singularities by constructing a
quasi-unitarized singularity corrected IAM amplitude.

\begin{figure}[p]
\centerline{\psfig{figure=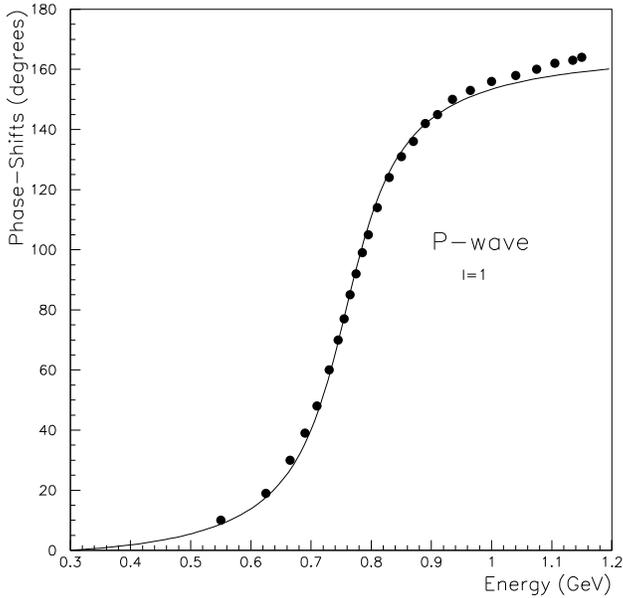,height=9.0cm}\hspace{0.5cm}
\psfig{figure=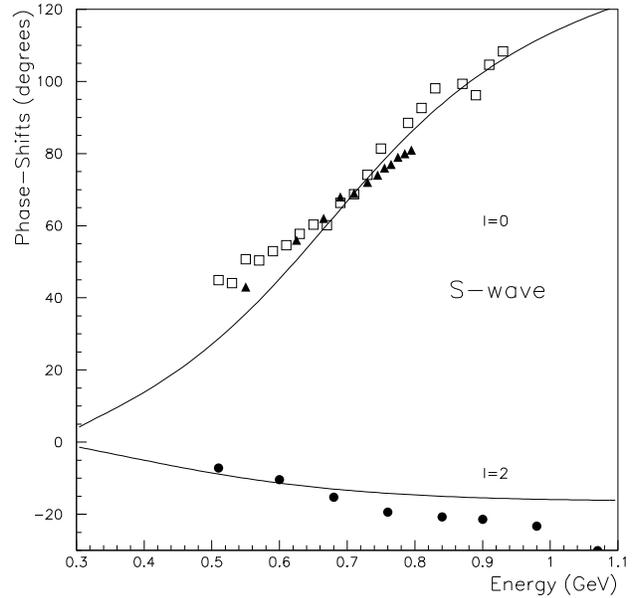,height=9.0cm}}
\caption{Results from fits of ChPT amplitudes 
to P- and S-wave phase-shifts, in
degrees, as functions of cms energy, in GeV. Same experimental data as
in Fig. \ref{iam_fits}.}
\label{upca_fits}
\end{figure}

For the sake of comparison, we show in Fig. \ref{upca_fits} the fits
of pure (crossing symmetric) ChPT 
amplitude to experimental phase-shifts, and evaluate, in turn, its
unitarity violation. The fits were done  as in
Ref. \cite{Bor}, using definition (\ref{upca}). 
P-wave fit presents the same quality as the
corresponding IAM one, while $I=0$ S-wave one is rather improved now. 
Here the parameters obtained are  $\lambda_1 =
0.007520$ and $\lambda_2 = -0.00653$, which have the
opposite signs in respect to those compatible with usual phenomenological
ones. It happens because, in order to reproduce $\rho$-resonance, a
complete inversion of the usual behaviour of the polynomial parts is
required. It is an illustration of the current understanding of why it
is not possible for pure ChPT amplitude to reach the resonance region.

\begin{figure}[p]
\centerline{\psfig{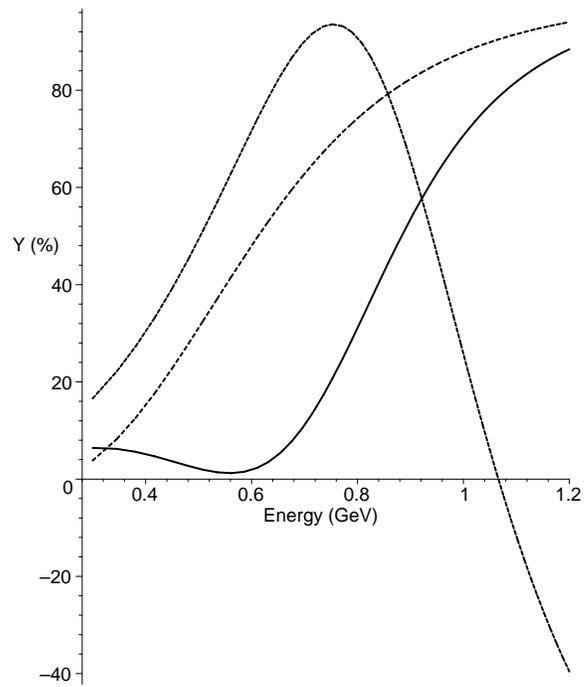}}
\caption{Unitarity violation (percentage) of ChPT amplitudes 
in P-wave (dotted), $I=0$ (solid) and 
$I=2$ (dashed) S-waves.} 
\label{upca_u}
\end{figure}

It is thus interesting to evaluate the unitarity violation of that
result. It is given, as before, by
$$ Y^{UPCA} = 
 \frac{ \mbox{Im}\,  {t_I}^{-1} - \rho}{
\mbox{Im}\,
{t_I}^{-1} + \rho} \, .$$
The results are shown in Fig. \ref{upca_u}. 
We obtained values that reach more than 90\%, for all three partial-waves. On
the other hand, all Roskies relations vanish, within the numerical
precision, of order $10^{-5}$, in this case. 

In summary  our results show that it is not possible for ChPT to exactly fulfill all
symmetry requirements, that is, by introducing elastic unitarity, a
lot of crossing symmetry is lost, as well as keeping the latter costs
a big amount of the former.

\begin{table*}
\begin{center} 
\begin{tabular}{|c||c|c|c|c|c|c|} 
\hline 
$i$ & $1$ & $2$ & $3$ & $4$ & $5$ &$6$ \\ \hline
IAM &   8.94\%&  0.66\%&  1.37\%& 1.00\%&  29.2\%&  37.3\% \\
\hline
\end{tabular}
\end{center}
\caption{Percentage deviations on Roskies relations 
$V_i$, according to Eqs. \ref{ints} and \ref{viola}.}
\end{table*}


\begin{thebibliography}{99}
\bibitem{Leu} J. Gasser and H. Leutwyler, \Journal{\NPB}{250}{1985}{465} 
\bibitem{Wei} S. Weinberg, \Journal \PRL {17} {1966} {616}
\bibitem{dob} A. Dobado, M. J. Herrero and T. N. Truong, \PLB
{\bf 235} (1990) 134; \\ A. Dobado and J. R. Pel\'aez,  
\PR {\bf D56} (1997) 3057.
\bibitem{Ro} R. Roskies, \Journal \NCA {65A}{1970}{467} 
\bibitem{Bor} J. S\'a Borges, J. Soares Barbosa and V. Oguri, \Journal
\PLB {393} {1997} {413}
 \bibitem{pro} S.D. Protopopescu {\em et al.}, {\em Phys. Rev.} {\bf
D7} (1973)  1279.
 \bibitem{estra} P. Estrabrooks and A.D. Martin,  {\em Nucl. Phys. B}
{\bf 79} (1974) 301. 
 \bibitem{lost} M.J. Lost {\em et al.},  {\em Nucl. Phys. B} {\bf 69}
(1974) 185. 
\end{thebibliography}
\end{document}